\documentclass[preprint,12pt,3p]{elsarticle}
\usepackage{epsfig,amssymb,amsfonts,amsmath,mathtools,bm,color}
\biboptions{sort&compress}

\newcommand{\X}{X(3872)}

\newcommand{\vel}{{\bm l}}

\newcommand{\veep}{{\bm \epsilon}}
\newcommand{\h}{\vphantom{$\int_0^1$}}
\newcommand{\Br}{{\rm Br}}

\journal{Physics Letters B}

\begin{document}

\begin{frontmatter}

\title{Heavy-quark spin symmetry partners of the $\X$ revisited}

\author[1,2]{V. Baru}
\author[1]{E. Epelbaum}
\author[1]{A. A. Filin}
\author[3]{C. Hanhart}
\author[3,4]{Ulf-G. Mei\ss ner}
\author[2,5,6]{A. V. Nefediev}

\address[1]{Institut f\"ur Theoretische Physik II, Ruhr-Universit\"at Bochum, D-44780 Bochum, Germany}
\address[2]{Institute for Theoretical and Experimental Physics, B. Cheremushkinskaya 25, 117218 Moscow, Russia}
\address[3]{Forschungszentrum J\"ulich, Institute for Advanced Simulation, Institut f\"ur Kernphysik and
J\"ulich Center for Hadron Physics, D-52425 J\"ulich, Germany}
\address[4]{Helmholtz-Institut f\"ur Strahlen- und Kernphysik and
Bethe Center for Theoretical Physics, Universit\"{a}t Bonn, D-53115 Bonn, Germany}
\address[5]{National Research Nuclear University MEPhI, 115409, Kashirskoe highway 31, Moscow, Russia}
\address[6]{Moscow Institute of Physics and Technology, 141700, 9 Institutsky lane, Dolgoprudny, Moscow Region, Russia}

\begin{abstract}
We  revisit the consequences of the heavy-quark spin symmetry for the possible spin partners of the $\X$. We 
 confirm that, if the $\X$ were a $D\bar{D}^*$ molecular state with the quantum numbers $J^{PC}=1^{++}$, then in 
the strict heavy-quark limit there should 
exist three more hadronic molecules degenerate with the $\X$, with the quantum numbers $0^{++}$, $1^{+-}$, and $2^{++}$ 
 in line with previous results reported in the literature. We demonstrate 
that this result is robust with respect to the inclusion of the one-pion exchange interaction between the $D$ mesons. However, this is true only
if all relevant partial waves as well as particle channels which are coupled via the pion-exchange potential
are taken into account. Otherwise, the heavy-quark 
symmetry is destroyed even in the heavy-quark limit. Finally, we solve the coupled-channel problem in 
the $2^{++}$ channel with nonperturbative pions beyond the heavy-quark limit 
and, contrary
to the findings of previous calculations with perturbative pions,
 find for the spin-2 partner of the $\X$ a significant shift of the mass as well as a width of the order of 50 MeV.
\end{abstract}


\end{frontmatter}

\section{Introduction}

In the previous decade, lots of states were found experimentally in the heavy quarkonium mass range that did
not at all fit into the scheme predicted by the until then very successful constituent quark model---for a review see, {\em e.g.}, 
Refs.~\cite{Brambilla:2010cs,Eidelman:2016qad}. Amongst those many states, the $\X$ is special not only because it was the first such an
extraordinary state discovered---it was first seen by the Belle Collaboration in 2003~\cite{Choi:2003ue}---but also because it resides extremely 
close to the $D^0\bar D^{*0}$ threshold. Indeed, with a mass $M_X=3871.68\pm0.17$ MeV \cite{Agashe:2014kda} its binding energy is as small as
\begin{eqnarray}
E_X=m_0+m_{*0}-M_X=0.12\pm 0.30~\mbox{MeV},
\label{EXexp}
\end{eqnarray}
where $m_0$ ($m_{*0}$) denotes the mass of the $D^0$ ($D^{*0}$) meson~\cite{Agashe:2014kda}. Thus it has been regarded as one of the most promising 
candidates for a hadronic molecule, which may be either an $S$-wave bound 
state~\cite{Voloshin:1976ap,DeRujula:1976zlg,Tornqvist:2004qy,Tornqvist:1991ks,Swanson:2003tb,Wong:2003xk} or a virtual
state in the $ D\bar D^*$ system~\cite{Hanhart:2007yq}; both possibilities are in line with its quantum numbers, which were determined by the
LHCb Collaboration to be $J^{PC}=1^{++}$~\cite{Aaij:2013zoa}. 
Other models exist in addition to the hadronic molecule interpretation, which include 
$\chi_{c1}(2P)$~\cite{Barnes:2005pb}---the first radial excitation of the $P$-wave charmonium $\chi_{c1}(1P)$,---a tetraquark~\cite{Faccini:2012pj}, 
a mixture of an ordinary charmonium and a hadronic
molecule~\cite{Close:2003sg,Suzuki:2005ha}, or a state generated in the coupled-channel dynamical scheme~\cite{Kalashnikova:2005ui,Danilkin:2010cc}.

One of the celebrated theoretical tools used in studies of hadronic states with heavy quarks is the Heavy-Quark Spin Symmetry (HQSS). HQSS is based 
on the observation that for $\Lambda_{\rm QCD}/m_Q\to 0$, with $m_Q$ denoting the quark mass, the strong 
interactions in the system are independent of the heavy quark spin. Then, although in case of the charm quark $\Lambda_{\rm QCD}/m_c\simeq 0.2$ is 
sizable and one expects non-negligible corrections to the strict symmetry limit,
constraints from HQSS can still provide a valuable guidance also in the charm sector and in particular
for the $\X$  \cite{Hidalgo-Duque:2013pva}. Meanwhile it was demonstrated in Ref.~\cite{Cleven:2015era} that the 
consequences of HQSS are very different for the 
different scenarios for the $X$. It is therefore crucial to refine the quantitative predictions for the
various scenarios. In this work we focus on a hypothesis that the $\X$ is a molecular state and investigate
the consequences that arise from HQSS as well as its leading violations. 
In particular, in Refs.~\cite{Bondar:2011ev,Voloshin:2011qa,Mehen:2011yh} the spin partners of the 
isovector states $Z_b^+(10610)$ and $Z_b^+(10650)$ were investigated in the molecular picture and several 
degenerate states were predicted. Similarly, it was argued in Refs.~\cite{Nieves:2012tt,Guo:2013sya} that 
one should expect a shallow bound state in the $D^*\bar D^*$ channel with the quantum numbers $J^{PC}=2^{++}$ --- the molecular partner of the \X. 
In Ref.~\cite{Albaladejo:2015dsa}, 
based on an effective field theory with perturbative pions (X-EFT), the width of this state was estimated to be as small as a few MeV.

In all mentioned studies as well as in this work an assumption is made for the dominant molecule
component of the wave functions of the states and observable implications of this assumption are investigated.
In reality one can expect that there is an admixture of different components too. However,
given the current quality of the data it appears unclear whether or not the effect of the subdominant components can be 
identified reliably within a given state. An exploratory study of the possible impact of genuine quarkonium states on 
the formation of the molecular spin multiplets is presented in Ref.~\cite{Cincioglu:2016fkm}. In the future it 
would certainly be of interest to combine the insights presented in this paper with the ideas of 
Ref.~\cite{Cincioglu:2016fkm}.\footnote{Note that according to Ref.~\cite{Hammer:2016prh} it might well be insufficient 
to include just a single quarkonium state in each channel.}

In this paper we refine further the implications of HQSS for the $\X$ and its partners in the molecular picture and critically re-examine the 
findings of the above mentioned papers. In particular, we investigate in detail the implications of HQSS for the 
spin partners of the $X(3872)$ with and without one pion exchange (OPE). We adapt the methods of Ref.~\cite{Mehen:2011yh} to isoscalar 
states in the presence of OPE, with a special emphasis on the renormalisation to leading order in the heavy quark expansion.
Furthermore, we go beyond the heavy-quark limit to demonstrate that the scales emerging in the coupled-channel approach 
due to the nonperturbative treatment of the pions generate a significant width of the $2^{++}$ spin partner of the $\X$ as well as a sizeable 
shift of its mass.

\section{Pionless theory---contact interactions only}

\subsection{Strict heavy-quark limit: Spin partners of the $\X$}

Although pions play 
 an important role in realistic calculations of the spin partners, as we shall demonstrate below,
it is instructive to start from 
a simple analytically solvable model with the only $S$-wave contact interactions. 
The methods applied in this section to the isoscalar states in the 
charmonium sector are similar to those used in Ref.~\cite{Mehen:2011yh} for isovector states in the bottomonium sector. 
In this subsection we discuss the results at leading order (LO) in the heavy-quark expansion which we call the strict HQSS limit. In this case, the 
masses of 
the $D$ and $D^*$ are identical. The corrections due to the finite $D^*$-$D$ mass splitting will be discussed in subsection~\ref{HQSSbreakingCT}.

The basis 
states introduced in Ref.~\cite{Nieves:2012tt} read
\begin{eqnarray}
&&0^{++}:\quad\left\{D\bar{D}({^1S_0}),D^*\bar{D}^*({^1S_0})\right\},\nonumber\\
&&1^{+-}:\quad\left\{D\bar{D}^*({^3S_1},-),D^*\bar{D}^*({^3S_1})\right\},\nonumber\\[-3mm]
\\[-3mm]
&&1^{++}:\quad\left\{D\bar{D}^*({^3S_1},+)\right\},\nonumber\\
&&2^{++}:\quad\left\{D^*\bar{D}^*({^5S_2})\right\},\nonumber
\end{eqnarray}
where the individual partial waves are labelled as $^{2S+1}L_J$ with $S$, $L$, and $J$ denoting
the total spin, the angular momentum, and the total momentum of the two-meson system, respectively. We define the C-parity eigenstates as
\begin{equation}
D\bar{D}^*(\pm)=\frac{1}{\sqrt{2}}\left(D\bar{D}^*\pm D^*\bar{D}\right),
\end{equation}
which comply with the convention\footnote{
  Notice that a different convention for the C-parity operator was used in Ref.~\cite{Nieves:2012tt}. As a 
consequence, the off-diagonal transitions 
 of $V_{\rm LO}^{(0{++})}$ in Ref.~\cite{Nieves:2012tt} have different sign as compared to Eq.~(\ref{C0++}), see also 
Sec. VI A in Ref.~\cite{Guo:2016bjq} for further 
 details of our convention.}
 for the C-parity transformation $\hat{C}{\cal M}=\bar{\cal M}$. 

In this basis and for a given set of quantum numbers $\{JPC\}$, the leading-order EFT potentials $V^{(JPC)}_{\rm LO}$, 
which respect heavy-quark 
spin symmetry, read \cite{AlFiky:2005jd,Nieves:2012tt,Valderrama:2012jv}
\begin{eqnarray}\label{C0++}
&&V_{\rm LO}^{(0{++})}=
\begin{pmatrix}
C_{0a} & -\sqrt{3}C_{0b} \\
-\sqrt{3}C_{0b} & C_{0a}-2C_{0b}
\end{pmatrix},
\label{Vct0++}\\
&&V_{\rm LO}^{(1{+-})}=
\begin{pmatrix}
C_{0a}-C_{0b} & 2C_{0b} \\
2C_{0b} & C_{0a}-C_{0b}
\end{pmatrix}, 
\label{Vct1+-}\\
&&V_{\rm LO}^{(1{++})}=C_{0a}+C_{0b} \label{eq:contact2-a},\label{Vct1++} \\
&&V_{\rm LO}^{(2{++})}=C_{0a}+C_{0b} \label{eq:contact2-b},\label{Vct2++}
\end{eqnarray}
where $C_{0a}$ and $C_{0b}$ are two independent low-energy constants.

The generic matrix integral equation for the scattering amplitude $a^{(JPC)}(p,p')$ reads
\begin{equation}
a^{(JPC)}(p,p')=V^{(JPC)}(p,p')-\int dk\,k^2 \, V^{(JPC)}(p,k)G(k)a^{(JPC)}(k,p'), 
\label{CT:Eq:1++}
\end{equation}
and it simplifies considerably in the strict HQSS limit if only the leading-order contact interactions 
(\ref{Vct0++})-(\ref{Vct2++}) 
are included. Here $G(k)$ denotes the matrix of the propagators of the heavy meson-antimeson pair in the intermediate 
state.
In the single-channel case---see Eqs.~(\ref{Vct1++}) and (\ref{Vct2++})---it reads 
\begin{equation}
G(k)=\frac1{k^2/\bar{m}-E-i0}
\label{propsym}
\end{equation}
while for coupled channels---see Eqs.~(\ref{Vct0++}) and (\ref{Vct1+-})---$G(k)$ is a $2\times 2$ diagonal matrix with 
both nonzero elements given by 
Eq.~(\ref{propsym}). Here we used that in the strict HQSS limit the $D^*$- and $D$-meson masses $m_*$ and $m$, 
respectively,
coincide, $\bar m=m_*=m$.
For the quantum numbers $1^{++}$ and $2^{++}$ Eq.~(\ref{CT:Eq:1++}) reduces to a single equation with the solution
\begin{eqnarray}
a^{-1}=C_0^{-1}+\bar{m}\int dk \frac{k^2}{k^2 - \bar{m}E-i0},
\label{singlechannel}
\end{eqnarray}
where $C_0 = C_{0a}+C_{0b}$.
The poles appear at the energies where the inverse amplitude, $a^{-1}$, vanishes. Accordingly, the value of the 
low-energy constant $C_0$ can be 
fixed from 
the binding energy of the $\X$ (denoted below as $E_X$), used as input. Conversely, the binding energy in the $2^{++}$ 
channel, $E_{X_2}$, can be 
extracted 
from this 
equation, given that $C_0$ is known. Clearly, $E_{X_2}=E_X$ in the strict HQSS limit. 

 As shown in Refs.~\cite{Hidalgo-Duque:2013pva,HidalgoDuque:2012pq,Bondar:2011ev,Voloshin:2011qa,Mehen:2011yh}, in 
the heavy-quark limit, one can predict more states with the same binding 
energy. To this end, one can apply a unitary transformation \cite{Mehen:2011yh}, defined as
\begin{eqnarray}
U=
\left(
\begin{array}{cc}
\cos \phi& \sin \phi \\
- \sin \phi & \cos \phi \\
\end{array}
\right),
\label{udef}
\end{eqnarray}
to the matrix bound-state Eq.~(\ref{CT:Eq:1++}) for the $0^{++}$ and $1^{+-}$ states. It is easy to see then that, 
taking 
$ {\phi=-\pi/6}$ and $\phi=\pi/4$ for the $0^{++}$ and $1^{+-}$ potentials defined in Eqs.~(\ref{Vct0++}) and 
(\ref{Vct1+-}),
respectively, one arrives for both quantum numbers at the diagonalised potential
\begin{eqnarray}
\tilde{V}^{(JPC)}=UV^{(JPC)}U^\dagger=
\begin{pmatrix}
C_0 & 0 \\
0 & C_0^{\prime}
\end{pmatrix},
\label{Vtilde}
\end{eqnarray}
where $C_0=C_{0a}+C_{0b}$ and $ {C_0^{\prime}}=C_{0a}-3C_{0b}$.
Therefore, the poles in both channels are now defined from the equation
\begin{eqnarray}
{\rm det}\left[1+\int dk\, k^2\,\tilde{V}^{(JPC)}G(k)\right]=0,
\label{DetCT}
\end{eqnarray}
where the propagator matrix is unchanged under rotations \eqref{udef}, $UG(k)U^\dagger={G}(k)$, since in the strict 
HQSS limit it is proportional to 
the unit matrix. Equation~(\ref{DetCT})
has two solutions corresponding to the two different linear combinations of the low-energy constants; one of them, 
$C_0$, is the same for all 
quantum numbers, including $1^{++}$ and $2^{++}$ considered before---see Eq.~(\ref{singlechannel}). 
Therefore the coupled-channel problem defined by Eqs.~(\ref{Vct0++})-(\ref{propsym}) in the strict HQSS limit splits 
into disentangled equations 
which possess two decoupled solutions,
\begin{equation}
E_X^{(0)}=E_{X_2}^{(0)}=E_{X_1}^{(0)}=E_{X_0}^{(0)}\quad\mbox{and}\quad E_{{X_0^{\prime}}}^{(0)}=E_{ 
{X_1^{\prime}}}^{(0)},
\label{HQSSlimit}
\end{equation}
where $E_{X_0}^{(0)}$, $E_{X_1}^{(0)}$, and $E_{X_2}^{(0)}$ stand for the binding energies of the spin-0, spin-1, and 
spin-2 partners of
the $\X$ in the strict heavy-quark limit, respectively, defined by the combination of the low-energy constants $C_0$ 
while 
$E_{ {X_0^{\prime}}}^{(0)}$ and $E_{ {X_1^{\prime}}}^{(0)}$
label the binding energies of the two additional partner states defined by the low-energy constant $C_0^\prime$---see 
Eq.~(\ref{Vtilde})---if the 
potential 
$C_0'$ is strong enough to bring about bound states.

In summary, in the strict HQSS limit the state $\X$ should have three degenerate spin partner states with the quantum 
numbers $0^{++}$, $1^{+-}$, and 
$2^{++}$, all of them being isoscalar states, like the $\X$ itself.
In addition, to these four degenerate states, there might exist two further states with the quantum numbers $0^{++}$ 
and $1^{+-}$ with a 
binding energy governed by the other combination 
of the low-energy constants, $C_0'$. These additional states cannot be predicted from the mass of the $\X$ and they 
require additional 
experimental input.   These findings are in line with those reported in Ref.~\cite{Hidalgo-Duque:2013pva}.

\subsection{Inclusion of HQSS breaking corrections}
\label{HQSSbreakingCT}

Corrections to the HQSS limit at leading order in $\Lambda_{\rm QCD}/m_c$ give rise to 
the known $D^*$-$D$ mass splitting. For convenience, we define the quantities
\begin{eqnarray}
\delta=m_*-m=141~\mbox{MeV},\quad \bar{m}=(3m_*+m)/4=1973~\mbox{MeV}
\label{delta}
\end{eqnarray}
and find for the reduced masses of the $D\bar{D}$, $D\bar{D}^*$, and $D^*\bar{D}^*$ pairs to leading order in $\delta$
\begin{equation}
2\mu=\bar{m}-\frac34\delta,\quad 2\mu_*=\bar{m}-\frac{\delta}{4},\quad 2\mu_{**}=\bar{m}+\frac14\delta ,
\end{equation}
respectively.

We start with the uncoupled channels corresponding to the quantum numbers $1^{++}$ and $2^{++}$. If, as explained 
above, the low-energy constant 
$C_0$ is fixed from the binding energy of the $\X$ ($cf$. Eq.~(\ref{singlechannel})), then for its $2^{++}$ partner we 
have the equation
\begin{eqnarray}
0=2\mu_*\int dk \frac{k^2}{k^2+\gamma_X^2-i0}-2\mu_{**}\int dk \frac{k^2}{k^2+\gamma_{X_2}^2-i0},
\label{singlechannel2}
\end{eqnarray}
where the binding momenta are related to the respective binding energies as
\begin{eqnarray}
\gamma_X^2=2\mu_* E_X,\quad\gamma_{X_2}^2=2\mu_{**} E_{X_2},
\end{eqnarray}
and the binding energies are now defined with respect to the relevant thresholds, namely
\begin{eqnarray}
E_X={m}+m_*-M_X,\quad E_{X_2}=2m_*-M_{X_2}.
\end{eqnarray}

The integrals in Eq.~(\ref{singlechannel2}) are linearly divergent and need to be regularised, for example, by a sharp 
cut-off in momentum, 
$k<\Lambda$. Then, dropping all terms which vanish in the limit $\Lambda\to\infty$ and retaining only the leading-order 
terms 
in $\delta$, we find that
\begin{eqnarray} 
\gamma_{X_2}=\left(1-\frac{\delta}{2\bar{m}}\right)\gamma_X+\frac{\delta}{\pi\bar{m}}\Lambda+\ldots,
\label{gammaX2}
\end{eqnarray}
where the ellipsis stands for the small corrections of the order $O(\gamma^2_X/\Lambda)$ and 
$O(\delta^2\Lambda/\bar{m}^2)$. Equation (\ref{gammaX2}) 
relates the binding momentum of the $X_2$ bound state to the binding momentum of the $\X$ where the latter is treated 
as input which fixes the 
strength of the contact potential $C_0$. We therefore see that at order $O(\delta)$ an additional counter term is 
necessary to render the 
result for the mass of the $J=2$ partner of the $\X$ well defined. 

The value of the counter term may be estimated by associating $\Lambda$ with the mass scale related to the pionic 
degrees of freedom which are integrated out in the contact theory. Alternatively one may argue that $\Lambda$ should be 
of order of the typical 
hadronic scale 
1 GeV. Accordingly, the cut-off-dependent 
term in Eq.~(\ref{gammaX2}) may be estimated to range between 3 and 23 MeV for $\Lambda=m_{\pi}$ and $\Lambda=1$ GeV, 
respectively. 
This uncertainty is to be compared with the value of the binding momentum $\gamma_X$. To estimate the latter in the 
isospin
limit
for the $D$-meson masses we consider two alternative assumptions: (i) the binding energy $E_X$ takes the value 
quoted in Eq.~(\ref{EXexp})---this gives $\gamma_X\approx 15$ MeV---and
(ii) the mass of the $X$ coincides with the experimental one such that $E_X=2\bar m-M_X\approx 4.2$ MeV,
which leads to $\gamma_X\approx 89$ MeV. We conclude therefore that 
from the effective theory with only $S$-wave contact interactions the $X_2$ state is expected to lie within a few MeV 
below the $D^*\bar{D}^*$ 
threshold.

It is straightforward to check that similar cut-off-dependent corrections induced by the $D^*$-$D$ mass difference 
appear in the channels with the 
quantum numbers $0^{++}$ and $1^{+-}$. 
In addition, as soon as the $D^*$-$D$ mass difference is considered, the propagator matrix is not proportional to the
unit matrix anymore and thus 
the product of the potential and the propagator cannot be diagonalised. 
As a result, the poles which appear for these two quantum numbers are determined by both low-energy constants $C_0$ and 
$C_0^{\prime}$---see 
Eq.~(\ref{Vtilde})---simultaneously. Accordingly, the binding energies in the $0^{++}$ and $1^{+-}$ channels are no 
longer equal to that of the 
$\X$, \emph{cf.} Eq.~(\ref{HQSSlimit}).
In order to proceed let us \emph{assume} that there exists a $1^{+-}$ bound state near the $D\bar{D}^*$ threshold, 
which we label as $X_1$. 
Then both low-energy constants $C_0$ and $C_0^{\prime}$ can be determined independently using the binding momenta 
$\gamma_{X_1}$ and $\gamma_{X}$ of 
the 
$X_1$ and the $\X$ as input.
As a consequence, the binding momenta of the other $1^{+-}$ and $0^{++}$ states can be predicted analytically from a 
coupled-channel approach. 
It should be stressed, however, that the role played by the coupled-channel effects 
depends on the interplay of the splitting $\delta$ and a typical binding energy of the spin partner states $E_B$. Had 
the 
relevant relation between the scales been $\delta\ll E_B$, then the binding energies discussed in the previous 
section---see 
Eq.~(\ref{HQSSlimit})---would have acquired only small corrections, perturbative in $\delta$.
However, in the realistic case the situation is opposite, $\delta\gg E_B$---see Eq.~(\ref{delta}) for the 
physical value of $\delta$---which calls for a different expansion for the coupled-channel equations. In particular, 
now $\sqrt{\bar{m}\delta}$ can be treated as a large parameter, and the expansion can be performed in powers of the 
small ratio 
$\gamma_B/\sqrt{\bar{m}\delta}$, where $\gamma_B$ is the binding momentum corresponding to the binding energy $E_B$ in 
the given channel 
\cite{Mehen:2011yh}.
For example, by an explicit calculation one arrives at
\begin{equation}
\gamma_{X_1'}=\left(1-\frac{\delta}{2\bar{m}}\right)\gamma_{X_1}+\frac{\Lambda \delta}{\pi\bar 
m}-\frac{(\gamma_{X_1}-\gamma_X)^2}{\sqrt{\bar m 
\delta}}+i\frac{(\gamma_{X_1}-\gamma_X)^2}{\sqrt{\bar m \delta}}+\ldots
\label{gammaXprime}
\end{equation}
for the binding momentum of the other $1^{+-}$ state, residing near the $D^*\bar{D}^*$ threshold and here referred to 
as $X_1'$.
This result is remarkable in two respects. First, states belonging to different HQSS multiplets---see 
Eq.~\ref{HQSSlimit}---are strongly mixed by the 
coupled-channel dynamics, 
so that the binding energy of the $1^{+-}$ $D^*\bar{D}^*$ state $X_1'$ depends now on both input parameters $\gamma_X$ 
and 
$\gamma_{X_1}$. Second, the binding momentum $\gamma_{X_1'}$ acquires an imaginary part and so does the binding 
momentum $\gamma_{X_0'}$ 
of the $X_0'$ state (the spin-0 state near the $D^*\bar{D}^*$ threshold).
This is 
a reflection of the fact that beyond the strict HQSS limit in the systems with the quantum numbers $0^{++}$ and 
$1^{+-}$ transitions 
$D^*\bar{D}^*\to D\bar{D}^{(*)}$ are possible due to coupled channels already in the pionless theory. It is important 
to notice that such 
imaginary parts are controlled by unitarity and therefore they are cut-off-independent to leading order---see 
Eq.~(\ref{gammaXprime}). 
The inclusion of the OPE interaction brings about additional 
partial waves in all channels and makes the transitions $D^*\bar{D}^*\to D\bar{D}^{(*)}$ possible for the quantum 
number $2^{++}$ as well, so that 
$\gamma_{X_2}$ becomes complex too---in other words, also the state $X_2$ acquires a finite width.
Meanwhile, as is demonstrated by the calculations described below, OPE does not spoil the property that the width of 
the spin-partner state shows 
only a rather mild cut-off dependence,  which makes it possible to treat the broadening of $\Gamma_{X_2}$ found in 
the calculations with nonperturbative pions as a reliable prediction of the approach.

\section{Contact plus OPE interactions}
 
It is often claimed that OPE plays a crucial role for the formation of the $\X$---the existence of the latter was even 
predicted based
on a model that contained OPE only~\cite{Tornqvist:1991ks}. We shall therefore investigate now the possible role of OPE 
from an effective field 
theory point of view. Since OPE in leading order is in line with HQSS, its inclusion does not destroy the multiplet 
structure discussed above. 
However, as 
we shall demonstrate below, this is only true if both coupled channels and $D$ waves are included properly. Before 
studying this
issue for the full, nonperturbative system, for illustrative purposes, we start with a discussion of the OPE 
contributions to one-loop 
order. This is sufficient to make the mentioned features apparent from the divergence structure of the amplitudes.

\subsection{Strict heavy-quark limit: Renormalisation to one loop}
\label{sec:cancel}

\begin{figure}
\centering
\includegraphics[width=13cm]{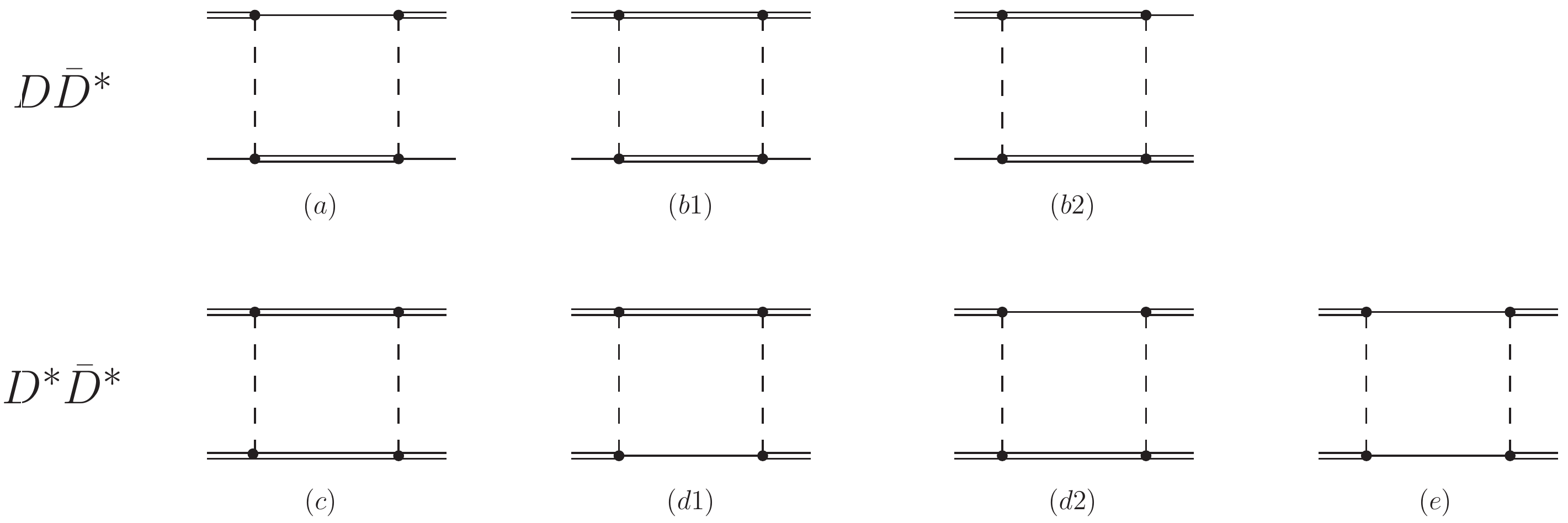}
\caption{One-loop diagrams which stem from two iterations of the OPE potential: The upper row shows contributions to 
the $D\bar{D}^*\to D \bar{D}^*$ 
transition potential and the lower row is for the $D^*\bar{D}^*\to D^*\bar{D}^*$ transition. Single (double) lines are 
for the $D$ ($D^*$) mesons and 
the dashed lines are for the pion.}
\label{fig:1loop}
\end{figure}

In this subsection we study the leading divergences of the one-loop diagrams which stem from two iterations of the OPE 
potential. We are going to demonstrate that, in the heavy-quark limit, the coefficients in front of the leading 
divergences in 
the $D\bar{D}^*\to D\bar{D}^*$ ($^3S_1$ partial wave) and $D^*\bar{D}^*\to D^*\bar{D}^*$ ($^5S_2$ partial wave) 
transition amplitudes coincide 
only if both $D\bar{D}^*$ and $D^*\bar{D}^*$ intermediate states are considered and all partial wave are kept in the 
calculation. The corresponding set of diagrams is shown in Fig.~\ref{fig:1loop}, where the upper row is for the 
$D\bar{D}^*\to D\bar{D}^*$ transition 
while the lower row is for the $D^*\bar{D}^*\to D^*\bar{D}^*$ transition. For convenience, we adopt the following 
convention: the meson floating 
along the upper line in each diagram is labelled by index 1 while the meson in the lower line is labelled by index 2. 
Also, particles in the final 
state are 
marked with a prime while particles in the intermediate state are marked with a double prime. 
 
In order to extract the leading divergences it is sufficient to retain only the loop momentum, denoted as $\vel$, in 
each vertex. Then, for example, 
the $D^*\to D\pi$ and $D^*\to D^*\pi$ vertices for the upper row read
\begin{eqnarray}
v^a(D^*\to D\pi)&=&\frac{g_c}{2 f_\pi}\tau_1^a(\veep_1\cdot\vel),\nonumber\\[-2mm]
\label{vDDpi}\\[-2mm]
v^a(D^*\to D^*\pi)&=&\frac{g_c}{2 f_\pi}\tau_1^a(-i[\veep_1\times\veep_1'] \cdot\vel),\nonumber
\end{eqnarray}
where  $\veep$  denotes   the polarisation vector of  the $D^*$  meson and $\tau^a$ is the isospin Pauli matrix. 
Further, $g_c=0.57$ is the dimensionless coupling constant which can be extracted from the $D^*\to D\pi$ width and 
$f_{\pi}=92.2$ MeV stands for the 
pion decay constant. 

The amplitudes $M_i$ ($i=a,\ldots e$) for the diagrams from Fig.~\ref{fig:1loop} can be schematically represented in 
the form 
\begin{equation}
M_i=C\int dl\; l^2(\hat{SL})_i (P_{\pi_1}G(l)P_{\pi_2}),
\label{Mi}
\end{equation}
where $C$ is a numerical coefficient, the same for all diagrams, $P_{\pi_i}$ ($i=1,2$) denote the pion propagators, and 
$G$ denotes as before the 
$D\bar{D}$, $D\bar{D}^*$ or $D^*\bar{D}^*$ time-ordered perturbation theory (TOPT) propagators, which are identical 
in all channels in the strict heavy-quark limit. 
The operator $(\hat{SL})_i$ labels the spin-orbit structure of the respective diagram. In particular, the leading 
divergences from 
diagrams $a$ and $b$ read
\begin{eqnarray}
(\hat{SL})_a&=&\int \frac{d\Omega_l}{4\pi}(\veep_1\cdot\vel)(\veep_1'\cdot\vel)(\veep_2''\cdot\vel)(\veep_2''\cdot\vel)
=\frac{1}{3}l^4(\veep_1\cdot\veep_1'),\\
(\hat{SL})_{b1}&=&\int\frac{d\Omega_l}{4\pi}(-i[\veep_1\times\veep_1'']\cdot\vel)(-i[\veep_1''\times\veep_1']
\cdot\vel)l^2
=\frac{2}{3}l^4(\veep_1\cdot\veep_1'),\\
(\hat{SL})_{b2}&=&\int\frac{d\Omega_l}{4\pi}(-i[\veep_1\times\veep_1'']
\cdot\vel)(\veep_1''\cdot\vel)(\veep_2''\cdot\vel)
(-i[\veep_2''\times\veep_2']\cdot\vel)=0, 
\label{div:b2}
\end{eqnarray}
where, as was explained above, prime (double prime) labels final (intermediate) particles. This yields
\begin{equation}
(\hat{SL})_{a,b}=(\hat{SL})_a+(\hat{SL})_{b1}+(\hat{SL})_{b2}=l^4(\veep_1\cdot\veep_1')
\label{div3s10}
\end{equation}
or, after projecting onto the ${}^3S_1$ partial waves, 
\begin{equation}
(\hat{SL})_{a,b}({}^3S_1)=l^4 P({}^3S_1)_i P^\dagger({}^3S_1)_i \ ,
\label{div3s1}
\end{equation}
where we used that the projection operator for the $^3S_1$ partial wave reads for a $D^*\bar D$ state 
\begin{equation}
P({}^3S_1)_i=\epsilon_{1i} \ .
\end{equation}
Similarly, for diagrams $c$-$e$ one gets
\begin{eqnarray} \nonumber
(\hat{SL})_{c}&=& \int \frac{d\Omega_l}{4\pi} (-i[\veep_1\times\veep_1''] \cdot\vel) (-i[\veep_1''\times\veep_1'] 
\cdot\vel) 
(-i[\veep_2\times\veep_2''] \cdot\vel) (-i[\veep_2''\times\veep_2'] \cdot\vel)\\
&=& \frac{l^4}{15}\, \left\{ 6 (\veep_1\cdot\veep_1') (\veep_2\cdot\veep_2') + (\veep_1\cdot\veep_2) 
(\veep_1'\cdot\veep_2')+(\veep_1\cdot\veep_2') 
(\veep_1'\cdot\veep_2)\right\},\\ \nonumber
(\hat{SL})_{d1}&=& \int \frac{d\Omega_l}{4\pi} (-i[\veep_1\times\veep_1''] \cdot\vel) (-i[\veep_1''\times\veep_1'] 
\cdot\vel) 
(\veep_2 \cdot\vel) (\veep_2' \cdot\vel) =\\
&=& \frac{l^4}{15}\, \left\{ 4 (\veep_1\cdot\veep_1') (\veep_2\cdot\veep_2') - (\veep_1\cdot\veep_2) 
(\veep_1'\cdot\veep_2')-(\veep_1\cdot\veep_2') 
(\veep_1'\cdot\veep_2)\right\},\\ \nonumber
(\hat{SL})_{e}&=& \int \frac{d\Omega_l}{4\pi} (\veep_1 \cdot\vel) (\veep_1' \cdot\vel) 
(\veep_2 \cdot\vel) (\veep_2' \cdot\vel) \\
&=& \frac{l^4}{15}\, \left\{ (\veep_1\cdot\veep_1') (\veep_2\cdot\veep_2') + (\veep_1\cdot\veep_2) 
(\veep_1'\cdot\veep_2')+(\veep_1\cdot\veep_2') 
(\veep_1'\cdot\veep_2)\right\},
\end{eqnarray}
and $(\hat{SL})_{d2}=(\hat{SL})_{d1}$, since the diagrams $d1$ and $d2$ differ from each 
other only by an index permutation for the intermediate 
particles. Summing up the individual contributions given above one arrives at
\begin{eqnarray}
(\hat{SL})_{c,d,e}=l^4(\veep_1\cdot\veep_1')(\veep_2\cdot\veep_2')
\end{eqnarray}
or, after projecting onto the $^5 S_2$ partial waves, 
\begin{eqnarray}
(\hat{SL})_{c,d,e}(^5S_2)=l^4P({}^5 S_2)_{ij}P({}^5 S_2) _{ij}^\dagger,
\label{div5s2}
\end{eqnarray}
where the ${}^5S_2$ projector has the form
\begin{equation}
P({}^5 
S_2)_{ij}=\frac12\left(\epsilon_{1i}\epsilon_{2j}+\epsilon_{1j}\epsilon_{2i}-\frac23\delta_{ij}
(\veep_1\cdot\veep_2)\right).
\end{equation}

\begin{table}[t]
\begin{center}
\begin{tabular}{|c|c|c|c|c|}
\hline
\h & & $D\bar{D}$ & $D\bar{D}^*$ & $D^*\bar{D}^*$\\
\hline
\hline
\h $1^{++}$ & $^{2S+1}L_J$ & --- & $^3S_1$\quad $^3D_1$ & $^5D_1$\\
\cline{2-5}
\h & Coeff. & --- & 1/9 \quad 2/9 & 2/3\\
\hline
\hline
\h $2^{++}$ & $^{2S+1}L_J$ & $^1D_2$ &$^3D_2$ &\hspace*{2mm}$^5S_2$\hspace*{5mm} $^1D_2$\hspace*{6mm} 
$^5D_2$\hspace*{3mm} 
$^5G_2$\\
\cline{2-5}
\h & Coeff. &2/15 &2/5 &1/9\hspace*{5mm} 2/45\quad 14/45\quad 0\\
\hline
\end{tabular}
\end{center}
\caption{The individual contributions to the coefficients (labelled as Coeff.) in front of the leading divergence in 
the $1^{++}$ 
$D\bar{D}^*(^3S_1)\to D\bar{D}^*(^3S_1)$ and $2^{++}$ $D^*\bar{D}^* (^5S_2)\to D^*\bar{D}^*(^5S_2)$ one-loop 
transitions from the 
intermediate $D\bar{D}, D\bar{D}^*$ and $D^*\bar{D}^*$ states in different (allowed) partial waves. Sum over all 
partial wave contributions is 
equal to 1 
in both channels in agreement with Eqs.~\eqref{div3s1} and \eqref{div5s2}.}\label{tab:waves}
\end{table}

By comparing the coefficients in front of the leading divergences in Eqs.~(\ref{div3s1}) and (\ref{div5s2}), one can 
see that they indeed coincide. 
This should not come as a surprise, given that, in the spin-symmetry limit, there is only one contact term available in 
the investigated 
transitions. It should be noted that, as soon as one of the external angular momenta is a $D$ wave, 
some momenta in the pion exchange amplitudes need to be identified with the external momenta in order to construct a 
$D$-wave projector. 
This reduces the degree of divergence of the corresponding integrals thus making them convergent. 

It is important to emphasise that the discussed equality of the leading divergences in the 
$D^*\bar{D}$ and $D^*\bar{D}^*$ channels---see Eqs.~(\ref{div3s1}) and (\ref{div5s2})---comes as a result of a delicate 
interplay between the 
contributions from different partial waves and different channels, as illustrated in Table \ref{tab:waves}.
For example, neglecting any $D$-wave
in the intermediate state destroys this equality, although $D$ waves still can be neglected altogether. 
Also, this equality is destroyed if any of the diagrams in Fig.~\ref{fig:1loop} is neglected (except for 
the diagram $b2$ which does not contribute to the leading divergence). In particular, in Ref.~\cite{Nieves:2012tt} OPE 
is included only for the 
diagonal transitions
$D\bar{D}^*\to D\bar{D}^*$ and $D^*\bar{D}^*\to D^*\bar{D}^*$ while the coupled-channel dynamics is neglected 
altogether. This implies that diagrams 
$b1$, $b2$, $d1$, $d2$, and $e$ are dropped in this work. However, this approximation leads to a violation of HQSS 
since the retained diagrams $a$ 
and $c$ have different coefficients in front of the leading divergence. Indeed, as can be seen from Table 
\ref{tab:waves}, neglecting 
the $D^*\bar{D}^*$ intermediate states in the $1^{++}$ channel leads to the coefficient 1/3 which is associated with 
diagram $a$
while neglecting the $DD$ and $D\bar{D}^*$ intermediate states in the $2^{++}$ channel results in the coefficient 7/15 
corresponding to diagram 
$c$. Hence, the single 
contact term present in the heavy-quark limit cannot absorb the divergences in the $1^{++}$ and $2^{++}$ channels 
simultaneously. As a 
consequence, the results of the pionfull calculations of Ref.~\cite{Nieves:2012tt} should reveal some cutoff 
dependence. 

One more comment on the sum over partial waves in the intermediate states is in order here. 
An explicit calculation in the partial wave basis shows
that diagram $b1$ in Fig.~\ref{fig:1loop} acquires a contribution from the intermediate $^3S_1$ partial wave which is, 
however, in contradiction 
with the required positive C-parity of the $D^*\bar{D}^*$ pair. Interestingly, the same contribution but with the 
opposite sign appears from 
diagram $b2$, although the net result from this diagram is zero---see Eq.~(\ref{div:b2}). This can be understood as 
follows: diagram $b1$ contains 
the sum of a contribution with positive C-parity and a contribution with negative C-parity while diagram $b2$ contains 
their difference.
Therefore the sum of diagrams $b1$ and $b2$ restores the required positive C-parity of the corresponding loop 
contribution while, at the same time, 
the 
UV-piece of diagram $b2$ vanishes since, in this limit, the contributions from different partial waves cancel.
This demonstrates that, although 
diagram $b2$ does not contribute to the UV-divergent piece of the one-loop amplitude, its omission has still to be done 
with caution to avoid 
problems with the C-parity of the amplitude. 

Notice that the power of divergence of the one-loop integrals for the diagrams in Fig.~\ref{fig:1loop} depends on the 
form of the $D^{(*)}\bar{D}^{(*)}$ propagator $G$. In this work we use nonrelativistic propagators, so that the 
one-loop contributions diverge 
linearly\footnote{One might be tempted to argue that in dimensional regularisation power divergences vanish. However, 
this is a 
scheme-dependent result which should be interpreted with caution, as discussed in detail in Ref.~\cite{Baru:2015nea}.} 
and higher powers of 
divergences show up starting from the third iteration of OPE. Then we choose the cutoff in the Lippmann-Schwinger-type 
equations of the order of a 
natural hard scale in the problem---see, for example, Refs.~\cite{Lepage:1997cs,Epelbaum:2006pt,Nogga:2005hy} in the 
context of nuclear EFT. 
Alternatively, if one uses a relativised propagator $G$, all iterations of OPE produce only logarithmic divergences 
which can be absorbed altogether 
by a single contact term for any value of the cutoff \cite{Baru:2015tfa}; see also Ref.~\cite{Epelbaum:2012ua} for the 
related work in the 
nucleon-nucleon problem. 
However, since the physical results should not depend on the particular method used, we here stick to the 
nonrelativistic propagator.

\subsection{Strict heavy-quark limit: nonperturbative inclusion of the OPE interactions}
\label{OPE}

We are now in the position to include the OPE interaction beyond one loop. Following the logic developed in the 
previous section, we start from the 
strict heavy-quark 
limit. Unlike the $S$-wave contact interactions, OPE allows for transitions to heavy-meson states in higher partial 
waves which have therefore to be included in an extended set of basis states,
\begin{eqnarray}\label{Eq:basis}
&&0^{++}:\quad\{D\bar{D}({}^1S_0),D^*\bar{D}^*({}^1S_0),D^*\bar{D}^*({}^5D_0)\},\nonumber\\
&&1^{+-}:\quad\{D\bar{D}^*({}^3S_1,-),D\bar{D}^*({}^3D_1,-),D^*\bar{D}^*({}^3S_1),D^*\bar{D}^*({}^3D_1)\},\nonumber\\[
-3mm]
\label{basisvec}\\[-3mm]
&&1^{++}:\quad\{D\bar{D}^*({}^3S_1,+),D\bar{D}^*({}^3D_1,+),D^*\bar{D}^*({}^5D_1)\},\nonumber\\
&&2^{++}:\quad\{D\bar{D}({}^1D_2),D\bar{D}^*({}^3D_2),D^*\bar{D}^*({}^5S_2),D^*\bar{D}^*({}^1D_2),D^*\bar{D}^*({}^5D_2),
D^*\bar{D}^*({}^5G_2)\},\nonumber
\end{eqnarray}
where, as before the C parity of the state is indicated explicitly in parenthesis whenever necessary.

The integral equations for the scattering amplitude can be written as
\begin{eqnarray}
a^{(JPC)}_{ik}(p,p')=V^{(JPC)}_{ik}(p,p')-\sum_j \int dk\,k^2 V^{(JPC)}_{ij}(p,k){G_j(k)}a^{(JPC)}_{jk}(k,p'),
\label{Eq:JPC}
\end{eqnarray}
where $i,j$ and $k$ label the basis vectors in the order they appear in Eq.~(\ref{Eq:basis}).
As before all propagators $G_j$ are equal in the heavy-quark limit. 
 
Performing a unitarity transformation on the basis states given in Eqs.~(\ref{Eq:basis}), one arrives at
\begin{eqnarray}
\tilde{a}^{(JPC)}_{ik}(p,p')=\tilde{V}^{(JPC)}_{ik}(p,p')-\sum_j\int dk\,k^2\tilde{V}_{ij}^{(JPC)}(p,k){G_j(k)}
\tilde{a}^{(JPC)}_{jk}(k,p'),
\end{eqnarray}
where $\tilde{a}^{(JPC)}=U^{(JPC)}a^{(JPC)}{U^{(JPC)}}^\dagger$ and 
$\tilde{V}^{(JPC)}=U^{(JPC)}V^{(JPC)}{U^{(JPC)}}^\dagger$. 
For a given set of quantum numbers $\{JPC\}$ one can find the operator $U^{(JPC)}$ 
such that the transformed potentials 
take a block-diagonal form (for the sake of transparency, the size of the blocks is quoted explicitly in parenthesis),
\begin{eqnarray}
&&\tilde V^{(0++)}(3\times 3)=A(2\times 2)\oplus B(1\times 1),\nonumber\\
&&\tilde V^{(1+-)}(4\times 4)=A(2\times 2)\oplus B(1\times 1)\oplus C(1\times 1),\nonumber\\[-3mm]
\label{Vtild}\\[-3mm]
&&\tilde V^{(1++)}(3\times 3)=A(2\times 2)\oplus D(1\times 1),\nonumber\\
&&\tilde V^{(2++)}(6\times 6)=A(2\times 2)\oplus D(1\times 1)\oplus E(3\times 3).\nonumber
\end{eqnarray}
The OPE contributes to all five submatrices, $A$, $B$, $C$, $D$, and $E$,
while the contact interaction contributes only to matrix $A$ (in the form of the linear combination 
$C_0=C_{0a}+C_{0b}$) and to matrix $B$
(as the linear combination ${C_0^\prime}=C_{0a}-3C_{0b}$). Accordingly, matrices $C$, $D$, and $E$ do not contain 
$S$-wave-to-$S$-wave transitions 
and 
are therefore quite unlikely to bring about bound states. Since matrix $A$ enters all four potentials in 
Eq.~(\ref{Vtild}) simultaneously, the 
degenerate bound states controlled by the contact potential $C_0$ appear in all four channels as before and, again as 
before, two 
additional 
degenerate bound states may exist in the channels $0^{++}$ and $1^{+-}$. They come from matrix $B$ and are controlled 
by the contact interaction 
$C_0^\prime$. 

\begin{figure}
\centering
\includegraphics[width=0.6\textwidth]{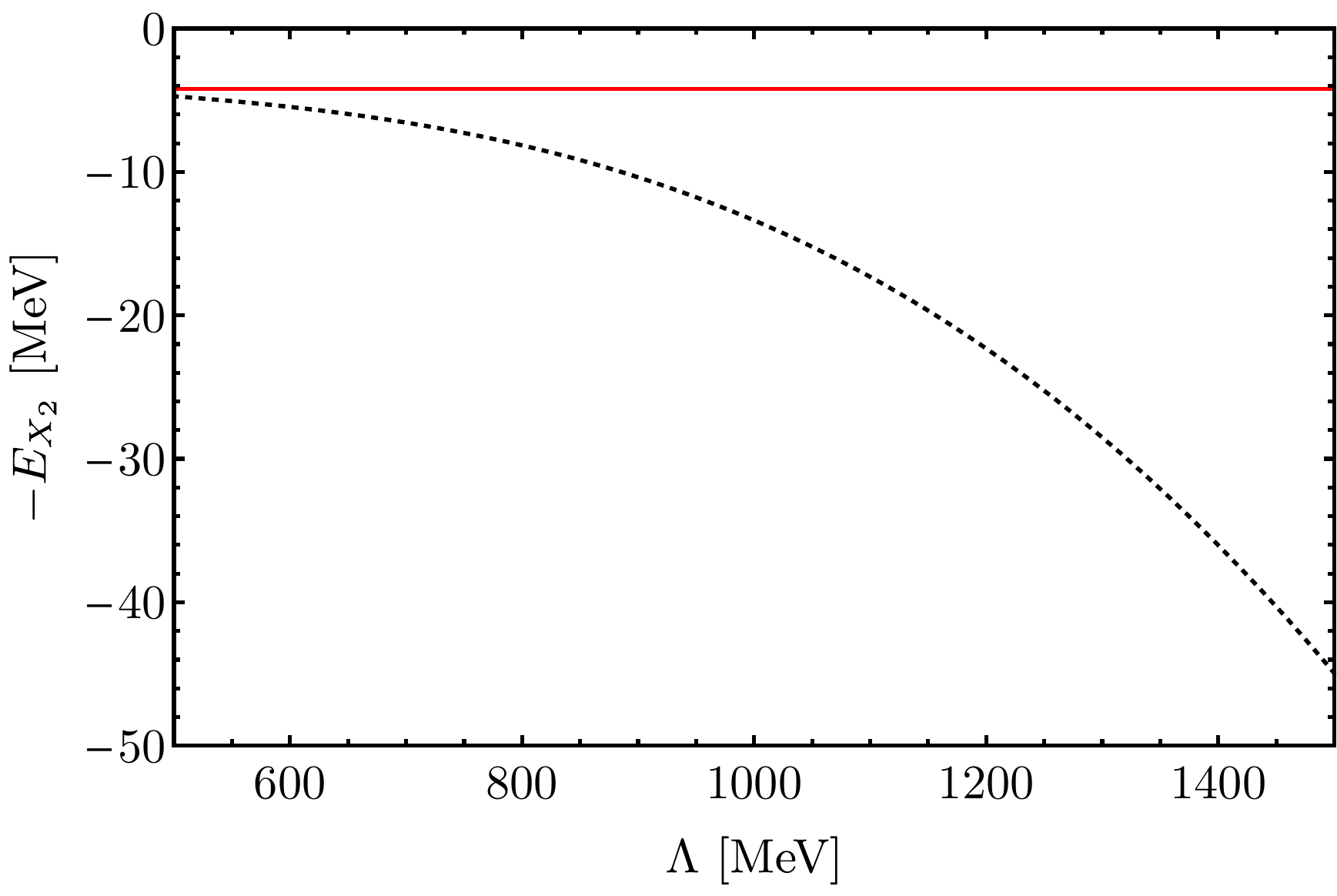}
\caption{The binding energy of the $2^{++}$ spin partner of the $\X$ in the strict heavy-quark limit: Red solid 
line---all coupled-channel 
transitions
are included, $E_{X_2}=E_X=4.2$ MeV; black dashed line---
particle coupled-channel transitions in the potentials 
$D\bar{D}^*\to D\bar{D}^*$ and $D^*\bar{D}^*\to D^*\bar{D}^*$
driven by the OPE interaction are neglected.
}
\label{fig:resHQS}
\end{figure}

We therefore observe that the specific pattern of degenerate bound states found in the purely contact theory survives 
the inclusion of the OPE 
interaction. Meanwhile, in line with the considerations of the previous subsection, we emphasise that the irreducible 
decomposition discussed above leads to degenerate states only if the basis vectors as given in Eqs.~(\ref{basisvec}) 
are included consistently. 
Specifically, 
the states remain degenerate if {\it all} $D$-wave-to-$D$-wave transitions are dropped in all coupled channels or/and 
if {\it all} 
$S$-$D$ transitions are dropped.\footnote{Note that transitions involving the $G$-wave contribute only to the matrix 
$E(3\times 3)$ and are therefore 
irrelevant for the formation of the discussed degenerate bound states in the heavy-quark limit.}
However, neglecting the particle coupled-channel 
dynamics or some higher partial waves immediately destroys the degeneracy deduced from the HQSS and leads to 
cutoff-dependent results for the 
partner 
states, since the cancellation of the divergences appears as a result of a delicate interplay between different partial 
wave amplitudes, as 
explained in the previous subsection. To illustrate this issue, in Fig.~\ref{fig:resHQS}, we show the cutoff dependence 
of the binding energy 
$E_{X_2}$ of the $2^{++}$ spin partner of the $\X$ in the strict heavy-quark limit. The arguments given above predict 
that $E_{X_2}$ takes 
exactly the same value as the binding energy of the $\X$ for which we stick to the value used in 
Ref.~\cite{Nieves:2012tt} for the isospin limit, namely $E_X=4.2$ MeV. In addition, $E_{X_2}$ should reveal no cutoff 
dependence. This is indeed the 
case for the full calculation---see the red solid line in Fig.~\ref{fig:resHQS}. On the other hand, neglecting the 
particle coupled-channel 
transitions in the potentials 
$D\bar{D}^*\to D\bar{D}^*$ and $D^*\bar{D}^*\to D^*\bar{D}^*$ governed by the OPE interaction results in strongly 
cutoff-dependent 
predictions---see the dashed black line in Fig.~\ref{fig:resHQS}. 
This approximation was used in 
Ref.~\cite{Nieves:2012tt} to predict the HQSS partner of the $\X$. Meanwhile, a quantitative comparison of the results 
contained in the 
aforementioned papers with those presented in Fig.~\ref{fig:resHQS} is not straightforward since (i) the result from 
Fig.~\ref{fig:resHQS} is 
obtained in the strict HQSS limit while in Ref.~\cite{Nieves:2012tt} the effects beyond the heavy-quark 
limit are also included and (ii) different regularisation schemes were used:
a rather soft exponential regulator of the form $\exp(-p^2/\Lambda^2)$ with $\Lambda=0.5$ and 1~GeV in 
Ref.~\cite{Nieves:2012tt} versus 
a sharp cutoff employed in this work.
 
\subsection{Beyond leading order}
 
As discussed above, the leading correction to the results obtained in the strict HQSS limit comes from
the $D^*$-$D$ mass difference that we denote as $\delta$---see Eq.~(\ref{delta}). The probably most spectacular new 
effect 
that 
comes into the system when both OPE and the leading corrections in $\delta$ are taken into account simultaneously is 
the finite width of the $2^{++}$ 
$D^*\bar{D}^*$ state that can now decay into the ${}^1D_2$ $D\bar{D}$ pair.  Within a theory with perturbative 
pions, in Ref.~\cite{Albaladejo:2015dsa}, this width was estimated to lie in the range from just a few units to about a dozen MeV 
depending on a particular model used for the pion form factor---see Table~I of Ref.~\cite{Albaladejo:2015dsa}.
Here we investigate for the first time the effect in a theory with nonperturbative pions. 

We expect the nonperturbative pion dynamics to be especially relevant for the transitions at hand since the 
momentum of the $D$ and $\bar{D}$ mesons in the final state that emerges from a shallow $D^*\bar{D}^*$ bound state is 
about 
\begin{equation}
q_1=\sqrt{2\mu(2\delta)}\approx 700~\mbox{MeV}.
\label{q1def}
\end{equation}
Transitions from the $D^*\bar{D}^*$ system to the $D\bar{D}^*$ final state also provide some inelasticity
and here the relevant momenta, of the order of
\begin{equation}
q_2=\sqrt{2\mu_*\delta}\approx 500~\mbox{MeV},
\label{q2def}
\end{equation}
are quite sizeable as well. In particular, both momenta are significantly larger than the pion mass.
A direct consequence of this is that $D$ waves fed by
the OPE are not subject to a kinematic suppression relative to the $S$ waves.

\begin{figure}
\centering
\includegraphics[width=0.4\textwidth]{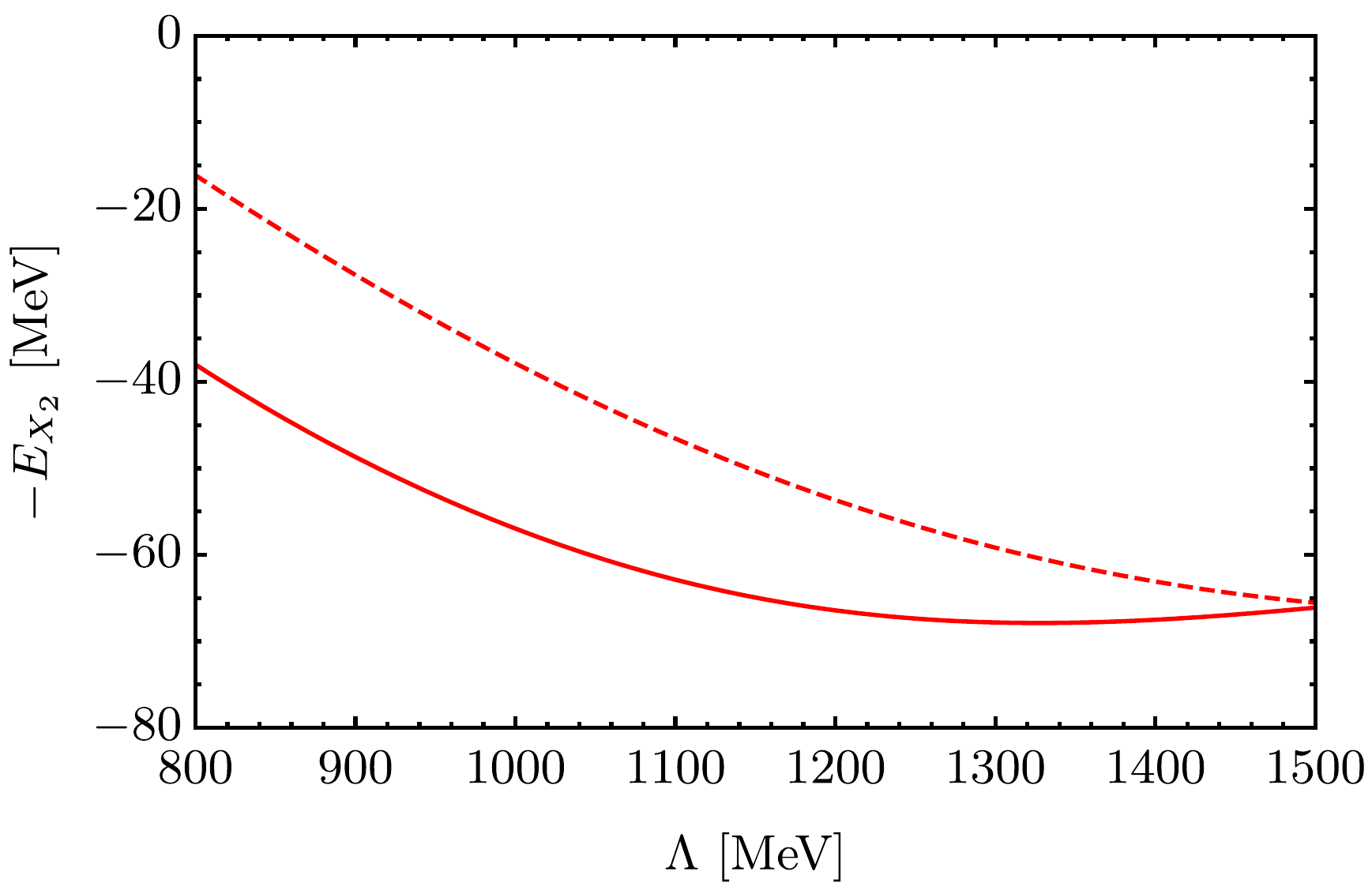}\hspace*{0.1\textwidth}
\includegraphics[width=0.4\textwidth]{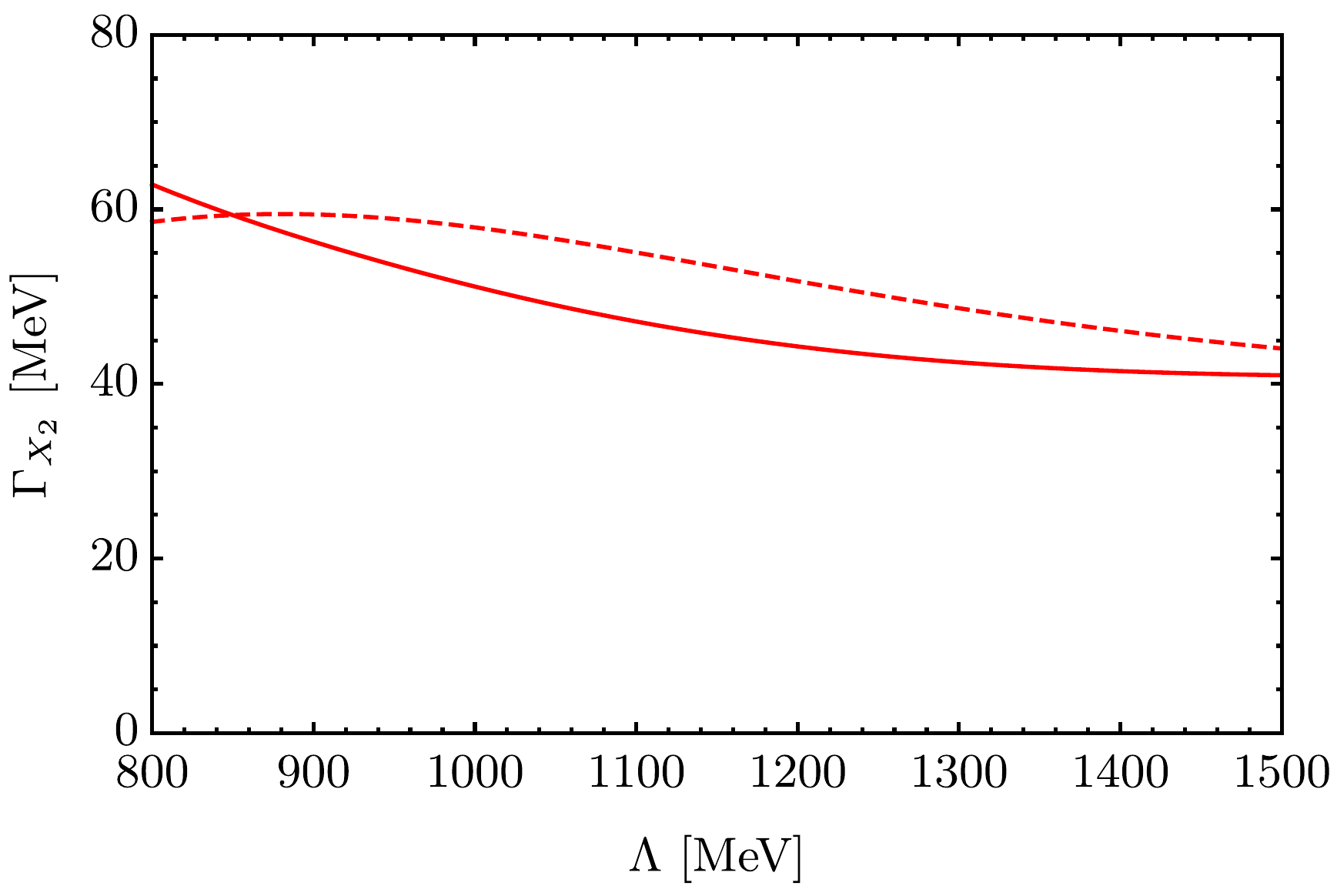}
\caption{The energy and the width of the $2^{++}$ bound $D^*\bar{D}^*$ state extracted using a Breit-Wigner 
parameterisation from the shape of 
the production rate, Eq.~\eqref{Eq:br}, as functions of the cutoff $\Lambda$ using two different regularisation schemes 
in the Lippmann-Schwinger 
equations : i)
sharp cutoff (solid lines), ii) the exponential regularisation of the form $f(p) = \exp(-p^6/\Lambda^6)$ (dashed 
lines).}
\label{fig:resX2}
\end{figure}

In order to calculate the observables, we proceed stepwise:
\begin{itemize}
\item Our leading-order potential consists of 
the low-energy constant $C_0$, adjusted to reproduce the $\X$ binding energy, and the static OPE potential 
(see Refs.~\cite{Valderrama:2012jv,Nieves:2012tt} and, in particular, Appendix~C of Ref.~\cite{Nieves:2012tt} 
for the explicit expressions which we reproduce). In order to connect to the results of 
Ref.~\cite{Nieves:2012tt} more directly, 
the $\X$ binding energy is chosen to be $E_X=4.2$~MeV.
\item The Green's functions $G_i$ ($i=D\bar{D}$, $D\bar{D}^*$, $D^*\bar{D}^*$) in Eq.~(\ref{Eq:JPC}) contain now the
 physical masses of the $D$ and 
$D^*$ mesons and in this way introduce into the system the intermediate momentum scales $q_1$ and $q_2$, defined in 
Eqs.~(\ref{q1def}) and 
(\ref{q2def}),
respectively.
\item The differential production rate ${d\Br}/{dE}$, as a function of the energy $E$ counted relative to the 
$D^*\bar{D}^*$ threshold,
is calculated from the convolution of the amplitude with a pointlike source,
\begin{equation}
\label{Eq:br}
\frac{d\Br}{d E}=\mbox{const}\times |J(E)|^2 k,\quad J(E)=\int dq\, q^2\frac{a_{D^*\bar{D}^*\to 
D\bar{D}}(q,k,E)}{E-q^2/m_*+i0}, 
\end{equation}
where $k=\sqrt{m(2\delta+E)}$ denotes the $D\bar D$ two-body phase space and $a_{D^*\bar{D}^*\to D\bar{D}}(q,k,E)$ 
denotes the solution
of the coupled channel scattering Eq.~(\ref{Eq:JPC}) in the half off-shell kinematics. 
\end{itemize} 
The function $J(E)$ has a clear Breit-Wigner shape that allows one to extract the (binding) energy and the width of the 
resonance from 
the shape of the below-threshold peak describing the $2^{++}$ $D^*\bar{D}^*$ bound state. 
These quantities are shown in Fig.~\ref{fig:resX2} as functions of the cutoff used to regularise the 
Lippmann-Schwinger equations. To assess the sensitivity of the results obtained to the form of the regulator we used 
two different regularisation 
schemes: the sharp cutoff $\theta(\Lambda-p)$ (the solid curves in Fig.~\ref{fig:resX2}) and the exponential function 
$\exp(-p^6/\Lambda^6)$ (the 
dashed curves in Fig.~\ref{fig:resX2}). 
Since we treat the momenta $q_1$ and $q_2$, defined in Eqs.~(\ref{q1def}) and (\ref{q2def}),
as soft scales, it is important to use a regulator that does not cut the momenta of their order. 
Both regulators mentioned above meet this criterion and lead to quite similar 
results for the parameters of the $X_2$ bound state, as seen from Fig.{~\ref{fig:resX2}}. 
The cutoff in the calculation is chosen to be of the order of the hard scale of the problem which 
is expected to be larger than $q_1$ but should not be taken too large to appropriately renormalise the scattering 
amplitude in the 
nonperturbative calculations \cite{Lepage:1997cs,Epelbaum:2006pt}. 
We therefore let the cutoff vary in the range 800-1500~MeV.   
Such a conservatively chosen  cutoff  range  allows us to estimate more reliably the impact of higher-order HQSS 
violating contact operators on the nonperturbative results. 
In particular, as will be seen below, the $\Lambda$-dependence of the results remains moderate 
even if one approaches larger values of the cutoff. For smaller cutoffs the separation of the soft and 
hard scales becomes worse and the results for the binding energy reveal larger dependence on the cutoff and on the 
choice of the regularisation scheme.

From the results presented in Fig.~\ref{fig:resX2} one is led to conclude that the scales emerging in the 
coupled-channel approach 
due to the nonperturbative treatment of the pions generate a significant shift of the $2^{++}$ spin partner of the $\X$ 
and make it as broad as 40-60~MeV. Those values are  a few times to an order of magnitude larger than predicted in 
Ref.~\cite{Albaladejo:2015dsa}.\footnote{We also checked by an 
explicit calculation that similar values of the parameters can be extracted from the differential rate of the two-step 
production process $D^*\bar{D}^*\to D\bar{D}^*\to D\bar{D}\pi$ from a pointlike source. } 
However, these findings have to be interpreted with caution. As was discussed in 
Subsec.~\ref{HQSSbreakingCT}, proceeding beyond the strict HQSS limit requires the presence of an extra counter term to 
absorb the cutoff dependence of the results. In the spirit of the numerical implementation of the renormalisation group 
equations, the residual 
$\Lambda$-dependence of the parameters of the $X_2$ bound state found for the cutoff varied in a reasonable range---see 
Fig.~\ref{fig:resX2}---provides a rough estimate of the 
importance of such a counter term. From the right plot in Fig.~\ref{fig:resX2} one can see that the observed dependence 
of $\Gamma_{X_2}$ on $\Lambda$ 
is quite mild  even when $\Lambda$ approaches the mass of
the $D$--meson,  where corrections to the heavy-quark limit could become significant. 
This appears to be in line with the discussion in the end of Subsec.~\ref{HQSSbreakingCT} 
where a similar 
observation was made for a purely contact theory beyond the HQSS limit. Therefore, the conclusion on the broadening of 
the $X_2$ state may be treated 
as a reliable
prediction of the approach used in this work. 
The discrepancy between this result and the conclusions of the previous study in
Ref.~\cite{Albaladejo:2015dsa} should be ascribed to the fact that in the latter work 
the $D$-wave contributions were suppressed by construction, since pions 
were included perturbatively. Contrary to this there is no suppression of the $D$ waves in our approach. 

 Meanwhile, the dependence of the binding energy $E_{X_2}$ on the value of the cutoff $\Lambda$ as well as on the 
regulator
 employed (left plot in Fig.~\ref{fig:resX2}) appears to be quite strong.
 In addition, the large momentum scales $q_1$ and $q_2$ call 
 for an extension 
of the model in order to incorporate further effects important for the problem. 
In particular, other members of the SU(3) pseudo-scalar octet and probably the 
vector mesons, whose masses are comparable with the relevant scales in the system at hand, should be included. 
In addition, three-body effects 
related to 
the $D\bar{D}\pi$ dynamics may also play a role and should be included---see 
Refs.~\cite{Baru:2013rta,Baru:2011rs,Baru:2015tfa} for the earlier
works on the $\X$ using
nonperturbative pions and Refs.~\cite{Fleming:2007rp,Jansen:2015lha} for the works including pions perturbatively.
Therefore, while the results of 
our calculations indicate that the $X_2$ state is shifted downwards in mass as soon as the leading
HQSS violating effects are included, we are not able at present to quantify this effect reliably. 
However, as argued above, our estimate for the width of the $X_2$ in the range $ 50\pm 10$ MeV
from nonperturbative pions remains to be a stable prediction of our approach
as the variation of the width with the cutoff is small compared 
to the width itself---see the right plot in Fig.~\ref{fig:resX2}.
 
\section{Conclusions and outlook}

In this paper we investigated the role of the pion exchange interactions for the formation of the spin partners of the 
molecular state $\X$. We 
demonstrated explicitly that the inclusion of the OPE interactions does not spoil the results of the pure contact 
theory in the strict 
heavy-quark limit which predicts the existence of 3 degenerate spin partners of the $1^{++}$ state $\X$ with the 
quantum numbers
$0^{++}$, $1^{+-}$ and $2^{++}$. However, we found that this conclusion as well as 
other predictions of the effective field theory incorporating both the contact and the OPE interactions can be regarded 
as reliable if, and only 
if, all particle coupled-channel effects and all relevant partial waves for the pion-exchange potential are taken into 
account. We 
demonstrated analytically to one-loop order that any omission of these requirements results in a violation of the 
heavy-quark spin symmetry.
We further confirmed this observation by an explicit nonperturbative numerical calculation of the $\X$ spin-2 partner 
binding energy $E_{X_2}$ in the 
strict heavy-quark limit: once the relevant low-energy constant is fixed to reproduce the mass of the $\X$ for any 
given value of the
cutoff $E_{X_2}$ turns out to be independent of $\Lambda$ in the full model. 
On the contrary, neglecting the $D\bar D$ and/or $D\bar D^*$ coupled-channel effects (in the $D$ waves) we find a 
strong cutoff dependence of 
$E_{X_2}$ even in the strict HQSS limit. 

Proceeding beyond the HQSS limit brings the scale $\delta$---the $D^*$-$D$ mass difference. This
results in new effects caused by the coupled-channel dynamics. 
In particular, in case of the spin partners with the quantum numbers $0^{++}$ and $1^{+-}$ the spin-symmetry-violating 
terms in the 
heavy meson-antimeson propagators lift the
degeneracy argued for in the symmetry limit and make each pole sensitive to the strength of both leading-order 
low-energy constants individually
and not only to their sum which may be fixed from the mass of the $\X$.

In addition, we observe that, even without coupled channels, already the leading spin-symmetry violating contribution 
calls for an additional counter term for the $D^{(*)}\bar D^{(*)}$ scattering system in order to absorb the dependence 
of the results on the 
regulator. This might put into question the possibility of an accurate 
prediction of the spin partners of the $\X$.
We demonstrate by an explicit calculation that it is still possible to at least estimate
both the binding energy and the width for the spin partner of the $\X$ with the quantum numbers $2^{++}$. 
For this we performed a coupled-channel analysis of the $D^*\bar{D}^*$ state with these
quantum numbers and found that the coupled-channel effects in the effective field theory incorporating both the contact 
and the OPE interactions had 
a strong impact on the parameters of this state and resulted in a sizable shift of the corresponding pole of the 
scattering matrix. 
In particular, we found that the binding energy and the width of this spin-2 partner of the $\X$ 
both appeared  to be of the order of several dozens MeV, that is significantly larger compared to the values found 
in the literature. 
We argue that, while the increase of the $X_2$ binding energy can only be viewed as a qualitative result 
the conclusion on the broadening of the $X_2$ is related to unitarity and therefore is a reliable prediction of our 
approach.

We emphasise that further progress and the 
possibility of more accurate predictions for the partner states should rely on a study of the convergence pattern 
of the approach used and in particular on an estimate of the role of higher-order contact interactions with two 
derivatives. Although these 
terms are formally suppressed in chiral EFT they might appear relevant here due to the relatively large momenta 
involved in the problem---see
Eqs.~(\ref{q1def}) and (\ref{q2def}). In addition, a more sophisticated study should 
include the three-body scales related to the $D\bar{D}\pi$ dynamics and an investigation of the role of the other 
members of the SU(3) pseudoscalar 
octet and vector mesons, whose masses are comparable with the scales relevant for the problem.

\medskip

 We are grateful to F.-K. Guo,  J. Nieves and M.~P.~Valderrama for useful comments on the manuscript. 
This work is supported in part by the DFG and the NSFC through funds provided to the Sino-German CRC 110 ``Symmetries 
and the Emergence of Structure
in QCD'' (NSFC Grant No. 11261130311). A. N. acknowledges support from the Russian Science Foundation (Grant No. 
15-12-30014). Work of V. B. is supported by the DFG (Grant No. GZ: BA 5443/1-1).  
The work of UGM was also supported by the CAS President's International Fellowship Initiative (PIFI) (Grant No. 2015VMA076).

\end{document}